\documentclass[fleqn,usenatbib,onecolumn]{mnras}

\usepackage{newtxtext,newtxmath}

\usepackage[T1]{fontenc}

\DeclareRobustCommand{\VAN}[3]{#2}
\let\VANthebibliography\thebibliography
\def\thebibliography{\DeclareRobustCommand{\VAN}[3]{##3}\VANthebibliography}

\usepackage{graphicx}	
\usepackage{amsmath}	
\usepackage{supertabular}

\title{
Pushchino Multibeam Pulsar Search. IX. Detection of a minute-long transient on the LPA antenna
}
\author[Anpilogov \& Tyul'bashev]{A. M. Anpilogov, $^{2}$
S. A. Tyul'bashev,$^{1}$\thanks{E-mail: serg@prao.ru (SAT)}
\\
	$^{1}$ P.N. Lebedev Physical Institute of the Russian Academy of Sciences, Astro Space Center, Pushchino Radio Astronomy Observatory,\\
	Radiotelescopnaya 1a, Moscow reg., Pushchino, 142290, Russia \\
	({E-mail: serg@prao.ru})\\
	$^{2}$ St. Petersburg State University, Saint Petersburg, Russia
}%

\date{2026}

\pubyear{2026}

\begin{document}
	\maketitle

\begin{abstract}
A transient (LPA~J0108+13) with repeated bursts was detected on the Large Phased Array (LPA) radio telescope at a central frequency of 110.4~MHz in the direction of the radio galaxy 3C~33. The flux density of bursts ranges from tens to hundreds of Jy, and the duration of the bursts is $\simeq 1^m-4^m$. In daily observations, the total duration of which at the location of the transient exceeds 200 hours in the observation interval 2013-2025, 6 bursts were found. The nature of the source could not be determined. We believe that a new type of transients has been discovered.

Keywords: transients, long period transient (LPT), low frequency observations, repeated bursts
\end{abstract}

\maketitle 

\section {Introduction}

Radio transients (hereinafter referred to as transients) are sources of burst radiation in the radio range. There are many different types of radio transients (see, for example, the Introduction to \citeauthor{Stewart2016}, \citeyear{Stewart2016}). In this paper, we will talk about strong (from hundreds of mJy) bursts with a duration of tens of seconds to minutes. 

In radio astronomy, short-time burst structures associated with intraday variability are known (see, for example, \citeauthor{Jauncey2001}, \citeyear{Jauncey2001}). They appear due to the passage of radiation through the interstellar medium. The duration of the flashes can be less than an hour. Such variability is easy to detect by conducting regular observations of compact radio sources over the course of days or weeks. Observations can take place in short sessions. The main thing is that the session duration should be longer than the characteristic time of variability, or the time between sessions should be shorter than the characteristic time scale of variability. At the same time, the starting point, when observations begin, does not matter. If the source is examined regularly, its variability will be revealed. The observed variability is associated with random processes, so it is possible to determine the characteristic time of variability, but no period is found in the observations.

Another example of transient phenomena is very short bursts lasting from fractions of a millisecond to tens of milliseconds, which are characteristic of pulsar pulses. In 2006, rotating radio transients (RRAT) were discovered, having a galactic nature \citeauthor{McLaughlin2006}, \citeyear{McLaughlin2006}. In 2007, fast radio bursts (FRBs) of an extragalactic nature were discovered \citeauthor{Lorimer2007}, \citeyear{Lorimer2007}. Both RRAT and FRB can have multiple outbursts when repeated observations confirm the discovery of a new source, as well as single outbursts when there is no independent confirmation of the discovery. An important sign of the discovery of a new short transient is the dispersion delay of the signal in the frequency channels. Another feature is the signal scattering observed at sources whose radiation travels a long distance in the interstellar and intergalactic medium. Since the pulse duration of short and millisecond pulsars can also range from fractions of a millisecond to tens or hundreds of milliseconds (see the ATNF catalog\footnote{https://www.atnf.csiro.au/research/pulsar/psrcat /}; \citeauthor{Manchester2005}, \citeyear{Manchester2005}), the methods of searching for such transients actually repeat the methods of searching for pulsars by individual pulses \citeauthor{Cordes2003}, \citeyear{Cordes2003}.

Over the past 20 years, there have been works by \citeauthor{Hyman2005}, \citeyear{Hyman2005}, \citeauthor{Stewart2016}, \citeyear{Stewart2016}, \citeauthor{Varghese2019}, \citeyear{Varghese2019}, \citeauthor{Hurley-Walker2022}, \citeyear{Hurley-Walker2022}, \citeauthor{Hurley-Walker2023}, \citeyear{Hurley-Walker2023}, \citeauthor{Caleb2024}, \citeyear{Caleb2024}, \citeauthor{Hurley-Walker2024}, \citeyear{Hurley-Walker2024}, \citeauthor{Dobie2024}, \citeyear{Dobie2024}, \citeauthor{deRuiter2025}, \citeyear{deRuiter2025}, \citeauthor{Lee2025}, \citeyear{Lee2025}, \citeauthor{Dong2024}, \citeyear{Dong2024}, \citeauthor{Bloot2025}, \citeyear{Bloot2025}, \citeauthor{Dong2025}, \citeyear{Dong2025}  that talk about the discovery of new types of transients. They are powerful (from hundreds of mJy to hundreds of Jy) radiation sources, the typical signal duration of which can be from a dozen seconds to minutes. Methods for searching for transients with a duration of $\sim 1$~minutes (hereinafter in the text we will use the terms transients or minute transients) began to be developed only in recent years. 

The difficulties in finding minute transients are due to the fact that a signal that appears once in the antenna recordings will be considered by radio astronomers as interference. If a pulse source does not have any characteristic features indicating its extraterrestrial nature (for example, the signal dispersion delay during multichannel observations), strong evidence is needed to substantiate its astrophysical nature. Otherwise, the found pulse signal should be considered interference. In general, the wide variety of interference visible in the raw data forces radio astronomers to be extremely careful. 

The first case of detection of a minute transient is, apparently, the source GCRT J1745-3009, observed on a VLA at a frequency of 330~MHz in the direction of the Galactic center \citeauthor{Hyman2005}, \citeyear{Hyman2005}, \citeauthor{Hyman2007}, \citeyear{Hyman2007}. The transient had pulses that fit into a period of 77~min.. The pulse duration was 2-10 ~min., the peak flux density ($S_p$) reached $\approx 1$~Jy. 

In the works \citeauthor{Stewart2016}, \citeyear{Stewart2016}, \citeauthor{Varghese2019}, \citeyear{Varghese2019}, \citeauthor{Hurley-Walker2022}, \citeyear{Hurley-Walker2022}, \citeauthor{Hurley-Walker2023}, \citeyear{Hurley-Walker2023}, \citeauthor{Caleb2024}, \citeyear{Caleb2024}, \citeauthor{Hurley-Walker2024}, \citeyear{Hurley-Walker2024}, \citeauthor{Dobie2024}, \citeyear{Dobie2024}, \citeauthor{deRuiter2025}, \citeyear{deRuiter2025}, \citeauthor{Lee2025}, \citeyear{Lee2025}, \citeauthor{Lee2025}, \citeyear{Lee2025}, \citeauthor{Bloot2025}, \citeyear{Bloot2025}, \citeauthor{Dong2025}, \citeyear{Dong2025} that followed through 15-20 years after the discovery of this transient, it was shown that both single and multiple events can occur. The duration of some of the observed pulses or details in the pulses may be less than a second, but their typical duration is about a minute. The periods defined for recurring bursts vary by 50 times, from 7 minutes (\citeauthor{Dong2024}, \citeyear{Dong2024}) to 6.45 hours (\citeauthor{Lee2025}, \citeyear{Lee2025}). The observed peak flux densities of the burst vary by 4 orders of magnitude: from one hundred mJy (\citeauthor{Caleb2024}, \citeyear{Caleb2024}, \citeauthor{deRuiter2025}, \citeyear{deRuiter2025}, \citeauthor{Dong2024}, \citeyear{Dong2024}) at frequencies of 135, 600, 887~MHz to tens Jy of (\citeauthor{Stewart2016}, \citeyear{Stewart2016}, \citeauthor{Hurley-Walker2022}, \citeyear{Hurley-Walker2022}) at frequencies of 60, 150~MHz, and even hundreds of Jy (\citeauthor{Varghese2019}, \citeyear{Varghese2019}) at the frequency is 34~MHz.

The nature of periodically emitting transients, also called long-period transients (LPT) or ultra-long-period transients (ULP), is usually associated by the authors of original papers with either magnetars or white dwarfs with periodic radio emission. The analysis conducted in \citeauthor{Rea2024}, \citeyear{Rea2024} suggests that it is difficult to make a choice between models of a slowly rotating neutron star or a white dwarf. None of the models explains all the observed characteristics (period, period derivative, flux density). The nature of the once-detected minute transients remains unclear.

It is most convenient to search for minute transients using maps obtained by the method of aperture synthesis. The maps are generated by integrating the signal at different time intervals. The transient on a series of such maps looks like a blinking dot. The disadvantage of this search method is the relatively high computational complexity of obtaining and analyzing a series of maps.

From an observer's point of view, all the described transients can be divided into two categories. The first category includes sources whose radiation has been detected only once, despite long-term repeated observations. The nature of these transients remains unknown, although they are similar to long-period transients in terms of flare duration and flux density. The second category includes sources that have been registered multiple times. Repeated detection of periodically emitted pulses is a reliable criterion for the discovery of a new transient.

In this paper, we report on the multiple detection of a new strong transient on the Large Phased Array (LPA) radio telescope of the Lebedev Physical Institute (LPI). The transient was discovered by chance during a summer internship by students of St. Petersburg State University. During the internship, the students evaluated the main characteristics of the LPA, the main instrument of the Pushchino Radio Astronomy Observatory (PRAO). When searching for sources with varying brightness (variable sources) based on sky maps constructed in temperature units, it was found that one of the calibration sources unexpectedly showed a strong increase in flux density. 

In the following sections, we talk about the used archival observations of the LPA LPI and their processing (section "Observations and processing"), the results obtained (section "Results"). In the last section ("Discussion and conclusion"), alternative for interpreting the results are considered.

\section{Observations and processing}\label{sec:observations}

\subsection{Antenna Characteristics}\label{sec:antenna}

The LPA LPI is a meridian instrument. All antenna beams are aligned in height in the same plane. The central frequency of the radio telescope ($\nu$) is 110.4~MHz (wavelength 2.72~m), the bandwidth is 2.5~MHz. Daily round-the-clock monitoring observations of the sky have been conducted since 2012, after the construction of an additional independent directional pattern (LPA-3) (\citeauthor{Shishov2016}, \citeyear{Shishov2016}). Currently, it includes 128 beams and covers declinations from $-9^\circ$ to $+55^\circ$. Observations are registered on three recorders. The beam size is $0.5^\circ\times 1^\circ$, the shape of the beam obeys the dependence $[\sin x /x]^2$. Until 2013, observations were carried out on 48 beams (on one recorder), and until 2020 - on 96 beams (on two recorders).

Data is registered from the recorder simultaneously in two time-frequency modes. In these modes, the full frequency band is divided into 32 or 6 frequency channels with a sampling time of 12.5 and 100 ms. The data is written in hourly pieces, but it is actually continuous. After the end of the next hour, the data recording file is closed and the next hour recording file is opened.

A calibration step (noise generator) with a temperature of 2400~K is used to calibrate the observations. Every 4 hours, the antenna is turned off, for about 15~s, and the step is recorded on the recorder in the form of "OFF-ON-OFF". The "OFF" mode means turning off the antenna and amplifiers, the "ON" mode means turning off the antenna, turning on the amplifiers and the noise generator. The accuracy of the temperature output by the noise generator is approximately 2-3\%.

The main tasks for the LPA-3 --- these are the "Space Weather" (\citeauthor{Shishov2016}, \citeyear{Shishov2016}) project, which uses a low-time-frequency recording mode, and the pulsar (\citeauthor{Tyulbashev2016}, \citeyear{Tyulbashev2016}) and transient (\citeauthor{Tyulbashev2018}, \citeyear{Tyulbashev2018}) search project, which uses a high-time-frequency recording mode. A low time-frequency resolution mode is used to study the current characteristics of LPA-3 (effective area, sensitivity).

\subsection{Detecting a transient}\label{discovery}

As mentioned in the Introduction, the transient was found by chance. To select the source used to evaluate the antenna characteristics, sky maps were built in temperature units (a calibration step was used). Discrete sources on the maps are visible as bright dots that stand out against the background of the Galaxy. At the strongest sources observed on LPA-3, side lobes are also visible, spreading along both declination and right ascension. An example of such a map is shown in Fig.~1. For educational purposes, LPA monitoring observations were available from August 1 to August 18, 2024, and a sky map was built for each day. Then the reference card (the median averaging of a series of cards) was subtracted from the current cards. For visual analysis, the sequences of the maps were animated (an MP4 video is available on our website\footnote{https://bsa-analytics.prao.ru/en/publications/}). The source 3C~33 (J0108+1320) became visible on the obtained anomaly maps, which increased its brightness twice (on August 14 and 17) in comparison with the surrounding sources.

\begin{figure*}
	\includegraphics[width=\linewidth]{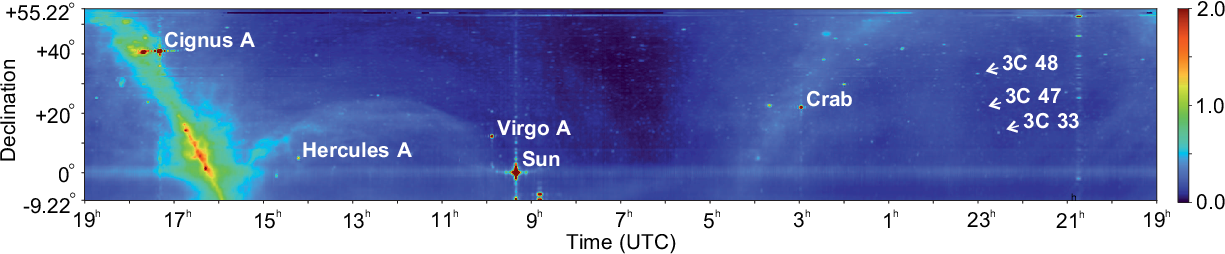}
	\caption{
Sky map for September 22, 2024 based on observations at LPA-3. Declinations are marked on the vertical axis, and UTC time is shown on the horizontal axis. Some bright radio sources have been signed. The flux density scale (the panel to the right of the panoramic image) is given in units of the calibration step. The colors from dark blue to burgundy show brightness temperatures from about 450~K to more than 3000~K. In the right part of the figure, source 3C~48 is marked (a powerful quasar clearly visible on the maps is a landmark for identifying weaker sources), as well as radio galaxies such as FRII 3C~47 and 3C~33, which fall on map fragments containing the studied transient for different dates and are presented in the following sections.
	}
	\label{fig:map_BSA3}
\end{figure*}

\subsection{Observation processing methodology}\label{sec:methods}

The observations were processed under the assumption that two strong sources entered the LPA beam simultaneously, one of which is the radio galaxy 3C~33, and the other is a probable galactic long-period transient. Other interpretations are discussed in the section "Discussion and conclusion".

Round-the-clock observations at the $-9^\circ<\delta < +42^\circ$ area has been going on for more than 10 years. To search for bursts, it was necessary to create a processing technique that would allow us to collect a 3C~33 light curve over the entire observation interval and identify "suspicious" days with increased flux density. At the next stage, it was necessary to check these days using the developed algorithms and visually. The main criterion for verification is a clear increase in the observed flux on the light curve and the absence of changes in the flux of the surrounding known radio sources.

The processing sequence included a number of obvious steps:\\
--- obtaining a 3C~33 light curve;\\
--- search for events with a noticeable increase in intensity compared to neighboring days;\\
--- verification of the reliability of the found ''suspicious'' events based on the constructed sky maps;\\
--- displaying hourly files with the original (raw) data, visual verification of the transient;\\
--- obtaining an estimate of the coordinates of the transient and estimating its flux density;\\
--- analysis of transient parameters;\\
--- evaluation of the period.

\subsubsection{Light curve}\label{sec:light_curve}

The light curve was extracted from the calibration source data with the pulse interference removed. A median filter with 5-second increments was used to remove interference. A calibration step was used to calibrate the signal in terms of temperature. The data was independently calibrated in all six frequency channels, after which the signals were added together.

All procedures were performed both for the main beam, where the source flux density is maximum, and for beams above and below. The 3C~33 source is located between two beams, so processing three beams allows you to determine the signal parameters, refine the apparent declination of the transient, and perform recording quality control.

The LPA-3 archive data processing program was written in Python using the astropy (\citeauthor{astropy2022}, \citeyear{astropy2022}) and astroplan (\citeauthor{astroplan2022}, \citeyear{astroplan2022}) libraries. Data was taken for processing from October 11, 2013 to July 31, 2025 (4324 sidereal days). The records for 278 days were missing or contained various kinds of interference and other errors that made processing impossible. Failures in the power supply or Internet connection to the server could lead to the absence of records. Thus, the data gaps amounted to 6.4\% of the light curve.

Some of the recordings involved in processing were affected by interference: lightning discharges, strong ionospheric storms, coronal mass ejections associated with $M$ and $X$ solar bursts, and others. Sometimes there is a redistribution of energy from the main lobe to the side lobes. There is no calibration step in the data part, which makes it impossible to equalize the gain across the frequency channels.

\begin{figure*}
	\includegraphics[width=0.6\linewidth]{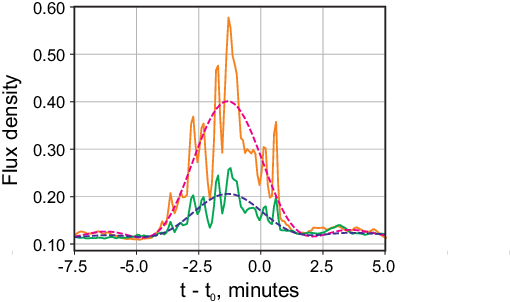}
	\caption{The figure shows the recording of the 3C~33 source for August 13, 2024 in two adjacent beams the day before the observed flare. The data contains the source model, the sum of the base signal level and the function $[sin(x)/x]^2$ (see equations \ref{eq:model} and \ref{eq:psf}). Strong ionospheric scintillations lasting tens of seconds is visible on the source. During calm ionospheric conditions, the source signal repeats the model almost exactly. The source record in the main beam is shown in orange, the red dotted line is the inscribed source model, the green and blue lines are the entry in the neighboring beam and the inscribed model.
	}
	\label{fig:map_fragments}
\end{figure*}

The ionosphere can shift the source's coordinate in right ascension and declination, as well as cause interference. To reliably estimate the parameters of the source signal, taking into account these circumstances, the model fitting method was chosen. The model is described by four parameters responsible for the source transmission curve and the base signal level:

\begin{equation}
	\label{eq:model}
	S(t - t_1) = A \cdot F(t - t_1) + B \cdot (t - t_1) + C
\end{equation}
where 
$S(t)$ is the recorded signal,
$F(t)$ is the response to a point source,
$t_1$ is observed transit time of the source,
$A$ is amplitude (peak flux density of source),
$B$ and $C$ is linear model of the basic signal level.

The response to a point source is described by the equation of the LPA radiation pattern adjusted for the cosine of declination of the source, which is responsible for the time of passage of the source through the meridian:

\begin{equation}
	F(t - t_1) = \left( \frac{\sin(x)}{x} \right)^2;
	\qquad
	x = (t - t_1) \cdot \frac{\pi D}{\lambda} \cdot \cos(\delta),
	\label{eq:psf}
\end{equation}
where
$D$ is the width of the LPA antenna along the parallel (187~m),
$\lambda$ is average LPA wavelength (2.72~m),
$\delta$ is declination of the source.

In Fig.~\ref{fig:map_fragments} a record of the 3C~33 source in two beams is presented. The main beam is shown in orange, where the flux density is maximum, and the neighboring beam is shown in green. The dotted lines show the source model embedded in the data (see equations \ref{eq:model} and \ref{eq:psf}).

\begin{figure}
	\includegraphics[width=0.8\linewidth]{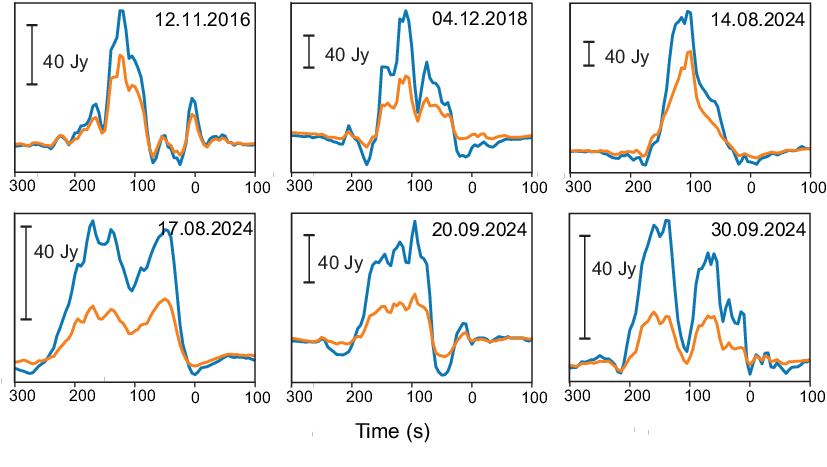}
	\caption{The transient is LPA~J0108+13 after subtracting 3C~33. The horizontal axis shows the time in seconds. A segment showing the flux density is shown along the vertical axis. The transient pulse in the main beam is shown in blue, and the adjacent beam is shown in orange. From the ratio of the amplitudes and based on the energy distribution between the beams according to the dependence $[sin(x)/x]^2$, the coordinate LPA~J0108+13 in declination was calculated. The transient profile sometimes becomes negative due to the contribution of the ionosphere to the amplification and attenuation of the 3C~33 flux before subtracting its transmission curve model.	
	}
	\label{fig:profile_13-08-2024}
\end{figure}

\subsubsection{Analysis of transit curves}\label{sec:transient_analisys}

The resulting light curve showed about hundreds of outliers, which were visually checked. The outliers were mainly related to various errors in the source data (for example, the absence of a calibration step). Automatic algorithms cannot be used to determine the characteristics of a transient. According to the works mentioned in the Introduction, the duration of bursts (pulses) It can differ up to 10 times from the same LPT. The shape of the pulse can also change. The apparent coordinate 3C~33 may shift from day to day in right ascension and declination due to the ionosphere. Therefore, the separation of radiation from 3C~33 and from the minute transient is a technical problem.

To solve this problem, we assumed that the source observed during the burst is determined by the convolution of the radiation pattern and the sum of the intensities from 3C~33 and the minute transient. The pulse shape, pulse duration, and minute transient flux density are unknown. The flux density of 3C~33 was considered unchanged. Since the angular sizes of 3C~33 are many times smaller than the size of the LPA beam, the apparent shape of 3C~33 is determined by the shape of the beam of the radiation pattern.  

It can also be expected that if the pulses of the observed transient were narrow (lasting up to tens of seconds), then against the background of a 3C~33 source, the total passage of which through the main lobe of the diagram is 5-7 minutes, a short burst (pulse) would be visible, standing out against the background of the 3C~33 source.

To estimate the transient parameters, the 3C~33 model is manually entered as a point source and then subtracted from the signal. Transient profiles during bursts after subtraction are shown in Fig.~\ref{fig:profile_13-08-2024}. The model amplitude of the signal was assumed to be close to the measured amplitude of the source in the previous and subsequent days from the event being tested. 

\begin{figure}
	\includegraphics[width=0.99\linewidth]{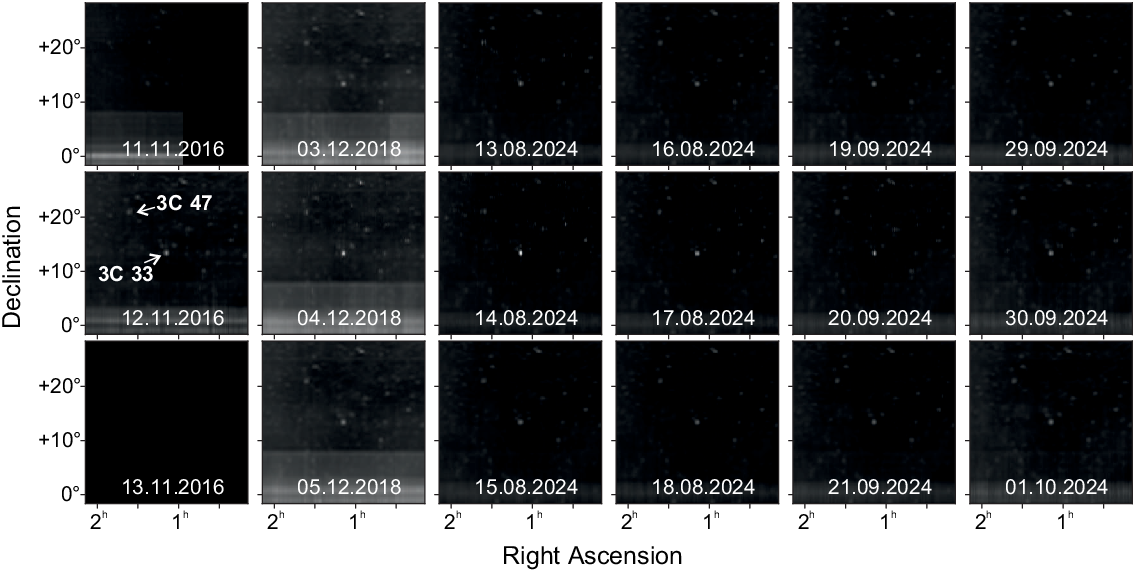}
	\caption{Fragments of sky maps. 3C~33 is centered, the coordinates are close to the coordinates of the ICRS system. The maximum pixel value corresponds to the height of the calibration step. Six series of map fragments are laid out along the horizontal axis for each found event with a transient. The series consists of three consecutive days, from top to bottom: a fragment of the map the day before the burst, the day with the burst, and the next day. The data for November 11 and 13, 2016 are of poor observation quality, and the gaps in the data are filled in with black. The arrows mark the sources 3C~33 and 3C~47.}
	\label{fig:LPT}
\end{figure}

For the selected days with bursts of radiation, fragments of panoramic sky maps were constructed (shown in Fig.~\ref{fig:LPT}), containing 3C~33 and its surroundings for three consecutive days. A comparison of the middle row of map fragments (the day with the burst) with the rows above and below (the day before the burst, the day after the burst) shows that the brightness of the 3C~33 source placed in the center of the map changes, while, for example, the brightness of the 3C~47 source located on the fragment above and to the left it remained unchanged. After the analysis, we are talking about detecting 6 bursts of radiation in the direction of 3C~33.

\section{Results}\label{sec:results}

\subsection{Coordinates of the transient}\label{sec:coordinates}

The coordinate of the minute transient in right ascension was determined on the assumption that we were observing the LPT, and its pulses could fall into any part of the LPA radiation pattern. Then the right ascension of the transient can be estimated by averaging the deviations of the "mass center" of the pulses from the inscribed profile 3C~33. The "mass center" of the pulses is defined as the weighted average value of the time coordinate over a short section of the signal. The weights correspond to the remainder of the signal after subtracting the 3C~33 model and the baseline. On average, for six bursts, no deviation was detected between the "mass center" of the signal remnants and the inscribed model 3C~33: $\alpha_{\text{LPT}} - \alpha_{\text{3C~33}} = -4\pm 8$~s.

The coordinate of the declination transient was also determined relative to the inscribed model 3C~33. Assuming a point source and a narrow directional pattern of individual beams, the signal distribution between the beams is described by a function of the form $[sin(x)/x]^2$, similar to the one given earlier for right ascension (Eq.\ref{eq:psf}). This allows us to fit 3C~33 into the initial data, subtract the model from the signals in two beams, and consider the distribution of the remaining energy over two beams. The declination of the remainder of the signal for each time point of the signal was found as the only solution to the problem of drawing a radiation pattern of a known type through two points. Only the declination of the flash signal has a physical meaning, so as an estimate of the declination, a weighted average of the time series of declinations found was calculated, with weights equal to the intensity of the signal. It was found that, on average, for six bursts, the transient declination is lower than declination 3C~33 by $0.03^\circ\pm 0.01^\circ$.

Thus, the coordinates of the transient are estimated: $\alpha_{2000}=01^h08^m49^s$, $\delta_{2000}=13^\circ 18^\prime$. Taking into account the sizes of the LPA radiation pattern, these coordinates are practically indistinguishable from the coordinates of 3C~33 ($\alpha_{2000}=01^h08^m52.86^s$, $\delta_{2000}=13^\circ20^\prime13.8^{\prime\prime}$).

\subsection{Search of period}\label{sec:period}

The observed transient may be LPT. We converted the moments of the "mass centers" impulses to Julian dates and obtained a time series of six time points. When trying to find the largest common divisor for the intervals between bursts, a period of $P = 7.978157$ hours was found. The check showed that the found period is exactly equal to a third of a sidereal day: 86164.09054 (seconds in sidereal days) / 3600 (seconds in an Earth hour)/ 3 = 7.97815653 sidereal hours.

As noted earlier, the LPA is a meridian instrument, so observations of any source are possible once a day. If bursts occur in the observed source at the time when it crosses the LPA radiation pattern, we will detect a period in sidereal days and fractions (1/2, 1/3, 1/4, and so on) of sidereal days. The coincidence of up to the 6th decimal place of the found "period" and a one third of the sidereal days indicates the randomness of the bursts in time.

\subsection{Pulse shape and width}\label{sec:pulse}

We attribute the remaining signal after subtracting 3C~33 to the transient. In Fig.~\ref{fig:profile_13-08-2024} shows all detected transient pulses. It can be seen that on November 12, 2016, December 4, 2018, August 17, 2024, and September 30, 2024, a complex pulse structure is observed. Checking the source data shows that the ionosphere has been noticeably distorting the source all these days. It is difficult to estimate the contribution of the ionosphere after subtracting 3C~33, but based on the record LPA~J0108+13 for August 14, 2024, we expect that the pulse has a simple one-component form. The width of the pulse, determined at half its height, varies 5 times: from 52~s (08/14/2024) to 277~s (08/17/2024).

\subsection{Flux density}\label{sec:flux_density}

Assuming that we are observing a long-period transient unrelated to the radiation of the radio galaxy 3C~33, it can be used as a calibration source. The estimation of the flux density of 3C~33 in terms of temperature was carried out using the light curve obtained earlier. Using the values of the 3C~33 flux density at 178 and 80 MHz given in the NED database\footnote{https://ned.ipac.caltech.edu /}, we obtain the value of the spectral index $\alpha=0.7$ ($S\sim\nu^{-\alpha}$). Recalculating the flux density by the frequency of observations $\nu=110.4$~MHz, we get $S_{110.4}=78$~Jy. 

Since we know the amplitude of the transient after subtracting 3C~33, we can estimate the value of the peak flux density LPA~J0108+13. The flux densities from the bursts vary from 57~Jy to 237~Jy, i.e. by a factor of 4. At the same time, the integral flux density (fluence is the total energy along the flash profile in terms of 1 second) varies within smaller limits: from 8390 Jy, from (November 12, 2016) to 17700 Jy from (August 14, 2024), 2 times.

Information on the detected bursts is available in the Table~\ref{tab:transients}. The first and second columns of the table show the moments of observations of a previously determined "mass center" pulse in Coordinated Universal Time (UTC) and in modified Julian dates (MJD). Columns 3-5 show the pulse widths at half height ($W_{0.5}$), their peak flux density, and fluence ($S_p; S_f$).

\begin{center}
	\begin{table*}
		\caption{Detected bursts from the transient}
		\begin{tabular}{|l|l|c|c|c|}
			\hline
			UTC                 & MJD         & $W_{0.5}$~(s) & $S_p$~(Jy)& $S_f$~(Jy\,$\cdot$\,s) \\
			\hline
			12.11.2016 19:08:08 & 57704.797318 & 53      & 90  & 8\,390  \\
			04.12.2018 17:43:48 & 58456.738753 & 67      & 150 & 10\,600 \\
			14.08.2024 01:06:17 & 60536.046027 & 52      & 237 & 17\,700 \\
			17.08.2024 00:53:58 & 60539.037478 & 277     & 60  & 12\,900 \\
			20.09.2024 22:36:40 & 60573.942129 & 99      & 102 & 9\,220  \\
			30.09.2024 21:57:18 & 60583.914792 & 106     & 57  & 9\,980  \\
			\hline
		\end{tabular}
		\label{tab:transients}
	\end{table*}
\end{center}

\subsection{Distance to the transient}\label{sec:distance}

The distance to the LPT in papers where such an estimate exists (for example, \citeauthor{Caleb2024}, \citeyear{Caleb2024}, \citeauthor{Hurley-Walker2024}, \cite{Hurley-Walker2024}, \citeauthor{Dong2024}, \citeyear{Dong2024} and others) is based on the observed signal dispersion delay in the frequency channels. That is, the transients showed narrow details in pulses lasting one second or less. In these details, it was possible to trace the shift of the signal in the frequency band associated with its passage in the interstellar medium. Knowing the dispersion measure (DM) and the model of the medium, it is possible to estimate the distance to the LPT. Using 32-channel data, we searched for narrow details in the found pulses. No such details were found, so the distance to LPA~J0108+13 could not be estimated.

\section{Discussion and conclusion}

The nature of the detected transient is unknown. In this paper, we conducted an analysis based on the assumption that the galactic source fell on the beam of LPA-3 in the direction of the radio galaxy 3C~33. Let's consider the interpretation of the obtained result.

Firstly, the detected bursts may be an internal or external interference. Secondly, the found source may be a new class of sources of extragalactic or galactic origin.

\subsection{Internal or external interference}\label{sec:interference}

The digital receiver that records the antenna signal is built on programmable logic integrated circuits (FPGAs) (\citeauthor{Tyulbashev2016}, \citeyear{Tyulbashev2016}). The incoming signal or the signal generated in the antenna paths themselves is amplified and gets to the FPGA. Each FPGA serves eight LPA beams. In fact, any external interference should fall into all 128 beams of the LPA, and internal interference into all eight beams formed by the FPGA (\citeauthor{Samodurov2022}, \citeyear{Samodurov2022}).

The signals we found always fall into the same two beams of the LPA. The signal is always stronger in one of the beams. There is not even a faint additional signals in the other beams of the LPA. Therefore, we dismiss the possibility of a interference nature of the bursts.

\subsection{A new type of transient}\label{sec:extragalactic_source}

The calculated coordinate LPA~J0108+13 practically does not differ from the coordinate of the radio galaxy 3C~33. Let's consider two possible options: the source may be associated with 3C~33 (have an extragalactic origin), or the source may be located in Milky Way.

According to the NED database\footnote{https://ned.ipac.caltech.edu /} information, 3C~33 is an extended FRII radio galaxy (\citeauthor{Owen1989}, \citeyear{Owen1989}). It has a redshift of $z\simeq 0.06$ and is located at a distance of $R\simeq 270$~Mpc. Three main components are observed in the radio range: two jets of similar brightness and a less bright central region (\citeauthor{Leahy1991}, \citeyear{Leahy1991}). The hot spots of the jets, which form the main stream of radio emission, are divided into $\approx 4^\prime$. For declination 3C~33, the width of the main lobe of the LPA radiation pattern along the celestial meridian (along the main direction of the jet axis) is $\approx 50^\prime$, which does not allow us to resolve the structure of the radio galaxy.

When looking at the central region, 3C~33 is a Type II Seyfert galaxy, narrow lines (NLRG) are observed. According to the unified AGN model, the "torus"\, of cold gas and dust is obscured by a bright accretion disk. Studies of the central structure using Gemini Multi-Object Spectrograph and Hubble Space Telescope have established the angle between the plane of the disk and the plane of the sky equal to $\approx 65^\circ$. It was also observed that inside 2.5 kpc there is a higher gas velocity ($\approx\pm 450$~km/s) and a lower metallicity ($12+\log(O/H)\approx 8.5-8.8$) than is usually observed in active galactic nuclei. The authors suggest that this gas was recently released as a result of an interaction that probably triggered the nuclear activity of \citeauthor{Couto2017}, \citeyear{Couto2017}.

That is, at 3C~33, no details are observed that could give a powerful burst. On the other hand, FRB bursts up in galaxies of different classes, and their energy is similar to that observed by us, with the exception that the duration of the burst of our transient is 4-5 orders of magnitude longer than the duration of a typical FRB. Assuming that the transient is located at 3C~33 and its radiation is isotropic, it is possible to estimate its pseudo-luminosity. Based on the observed flux density of 100~Jy, pulse width of 100~s, frequency of observations of 0.11~GHz, distance to the galaxy $2.7 \cdot 10^5$~kpc will get $7.29 \cdot 10^{12}$~Jy $\cdot$ kpc$^2$. It is impossible to exclude the extragalactic origin of the transient, especially given the coincidence of the coordinates 3C~33 and the transient. At the same time, it should be assumed that a new type of extragalactic powerful transients has been discovered.

It can be assumed that the transient is a galactic source that came into line of sight with the galaxy 3C~33. In this case, we can talk about an object that flashes regularly, but its flashes are not periodic. Then its pseudoluminosity can be on the order of $1-10^4$~Jy $\cdot$ kpc$^2$ depending on the distance (0.1; 1; 10 kpc). We have not been able to detect such objects in the literature, so it can be argued that a new class of strong minute transients with irregular (non-periodic) bursts has been discovered.

In studies of bursts of different durations, you can see a diagram on which the sources are placed based on the duration of the bursts and their energy (see, for example, \citeauthor{Pietka2015}, \citeyear{Pietka2015}). In Fig.~5, we placed the source LPA~J0108+13. It can be seen from the figure that for the case of both galactic and extragalactic origin, the transient falls into an area where there are no known types of sources. The figure also shows that its radiation is coherent.

\begin{figure}
	\includegraphics[width=0.9\linewidth]{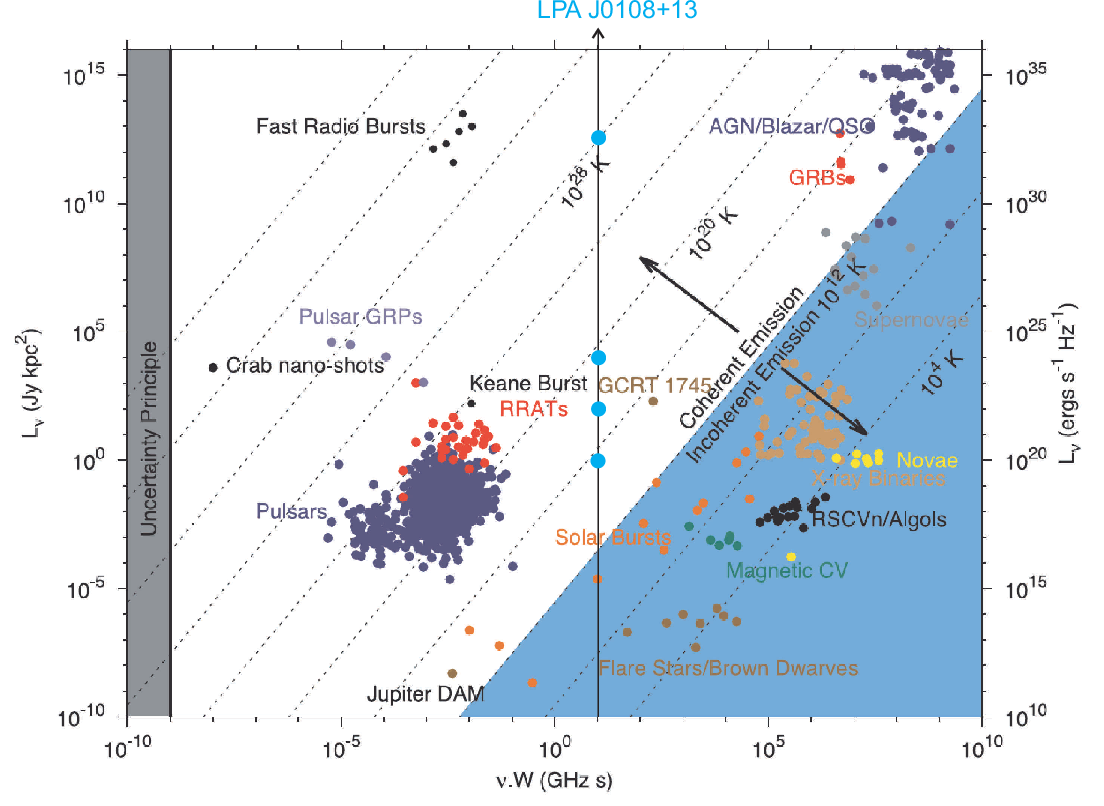} 
	\caption{
The transient LPA~J0108+13 in the luminosity diagram is the duration of the signal relative to known types of sources. The basis of the drawing is taken from the paper \citeauthor{Pietka2015}, \citeyear{Pietka2015}.    
	}
	\label{fig:diagramm}
\end{figure}

In about 200 hours of observations, 6 bursts were detected in the direction of 3C~33. That is, on average, over the entire observed interval, there is an burst every 33 hours. If we take 2024, bursts were observed on average every 5 hours. As can be seen from the Table~\ref{tab:transients}, the weakest bursts had a residual amplitude of at least 0.7 of the peak flux density of 3C~33. It is likely that with a decrease in the amplitude criterion of the remainder, the number of bursts will increase. Unfortunately, this is a difficult task for the LPA LPI radio telescope. We urge our colleagues to pay attention to this very active source and analysied its archived data.

In conclusion, we summarize the main result: We have discovered a new type of minute transients emitting powerful time-random bursts with coherent radiation. The nature of the source is not clear, it can have both galactic and extragalactic origin.

\section*{Data availability}

The source data is available upon request to the corresponding author.

\section*{Acknowledgements}

The authors would like to thank R.V. Baluyev for his valuable comments, as well as L.B. Potapova for her help with the design of the drawings and the paper.  SAT expresses its gratitude to the Russian Science Foundation for its support of the work (grant 22-12-00236-$\Pi$\footnote{https://rscf.ru/project/22-12-00236-$\Pi$/})

\end{document}